\documentclass[twocolumn]{aastex63}

\usepackage{acronym}
\usepackage{amsmath,amsfonts,amssymb}
\usepackage{mathrsfs}
\usepackage[utf8]{inputenc}

\usepackage[normalem]{ulem}
\usepackage{color}

\shorttitle{Neutrino Waveforms}
\shortauthors{Vartanyan et al.}

\begin{document}

\title{Gravitational Waves from Neutrino Emission Asymmetries in Core-Collapse Supernovae}
\correspondingauthor{David Vartanyan}
\email{dvartany@berkeley.edu}

\author[0000-0003-1938-9282]{David Vartanyan}
\affiliation{ Department of Physics and Astronomy, University of California, Berkeley, CA  94720}
\author{Adam Burrows}
\affiliation{Department of Astrophysical Sciences, Princeton, NJ 08544, USA}




\begin{abstract}
We present a broadband spectrum of gravitational waves from core-collapse supernovae (CCSNe) sourced by neutrino emission asymmetries for a series of full 3D simulations. The associated gravitational wave strain probes the long-term secular evolution of CCSNe and small-scale turbulent activity and provides insight into the geometry of the explosion. For non-exploding models, both the neutrino luminosity and the neutrino gravitational waveform will encode information about the spiral SASI. The neutrino memory will be detectable for a wide range of progenitor masses for a galactic event. Our results can be used to guide near-future decihertz and long-baseline gravitational-wave detection programs, including aLIGO, the Einstein Telescope, and DECIGO.
\end{abstract} 

\section{Introduction}
\label{sec:int}

Core-collapse supernova (CCSNe) explosions are dynamical events
involving extreme astrophysical conditions, with the central core reaching 
super-nuclear densities and temperatures of hundreds of billions Kelvin. 
The relevant timescales in CCSNe range from sub-milliseconds to seconds, the former
associated with the  dynamical bounce, rotation, and convective motions
and the latter due to secular evolution of the late-time neutrino emissions and 
explosion debris. No viable CCSNe explosion is spherical (\citealt{radice2017b,vartanyan2019,nagakura_res,burrows_2019,burrows_2019b,2019MNRAS.487.5304M}, with the possible exception of low-mass $\le$9.5 M$_{\odot}$ 
progenitors which have very steep density profiles and weakly-bound mantles, see e.g. \citealt{radice2017b}). 
In general, multi-dimensional effects are critical to both modeling a successful explosion itself, 
and parametrized turbulence in spherically-symmetric 1D models  
(\citealt{mabanta2019,2020ApJ...890..127C}) is less than adequate.
Due to time-changing quadrupolar motions, CCSNe are classic 
sources of gravitational waves (GWs), with most of the gravitational-wave 
energy ($\sim$10$^{-8}$ M$_{\odot}\mathrm{c}^2$, where M$_{\odot}$ 
indicates a solar mass) coming out at 100s to 1000s of Hertz.  A major 
discriminating spectral signature is the g-mode/f-mode of fundamental 
oscillation of the proto-neutron star (PNS) that evolves to higher 
frequencies as the core deleptonizes and cools on its way to the cold, 
catalyzed neutron star state.  This mode is excited by asymmetrically
infalling plumes of matter that hammer the PNS in the first seconds of 
core-collapse and explosion and is an important asteroseismological 
measure of core structure and early evolution.  The modern theory of 
this dominant GW component of core-collapse supernovae is summarized 
in \cite{vsg2018} and in \cite{radice2019}, and in references therein (see also \citealt{2017PhRvD..96f3005S,2018MNRAS.474.5272T,2019MNRAS.482.3967T,2020PhRvD.102b3027M}).

However, the low-frequency component below $\sim$10 Hz, not so easily measured 
even by next generation ground-based platforms, bears the stamp of 
important secular motions.  The first is due to ejecta motions themselves.
The explosions are generically asymmetrical, with matter ejection and core recoil
kicks on 0.1 to a few seconds timescales \citep{burrows1996,murphy2009,holgado_2019}.  
Interestingly, the metric perturbations do not necessarily return to zero, and the metric is permanently
shifted. This is akin to the classical  ``memory" effect \citep{christodolou1991,thorne1992}, 
but is due to asymmetrical matter ejection. The associated 
frequencies are $\sim$0.1 to 10 Hz. It is thought that pulsars are born with kicks
due either to asymmetrical ejecta or to asymmetrical neutrino emission and the 
associated momentum recoils.  Though momentum asymmetries are dipolar phenomena,
there is always an associated quadrupolar component \citep{vartanyan2019}.

Along with the longer-term secular matter component at low frequencies,
intriguingly there is a similar contribution due to asymmetrical neutrino
emissions \citep{epstein1978,turner1978,braginskii,burrows1996,emuller1997,kotake2005,kotake2009,2010JPhCS.229a2011K,kotake2011,emuller2012,2013ApJ...766...43M, Li_2018}. 
The neutrinos move at (very near) the speed of light, involve $\sim$0.1 to $\sim$0.3 solar masses equivalent, and are
generically emitted aspherically (see, e.g., \citealt{tamborra2013,vartanyan2019,walk2019}).  The shells of outgoing neutrinos
constitute time-changing quadrupoles that source gravitational radiation
at frequencies of $\sim$0.1 to 10 Hz \citep{burrows1996,emuller1997,emuller2012}. The high bulk 
mass-energy and speed of the neutrinos result in their almost complete domination
over the matter component for low-frequency gravitational-waves in the GW spectrum
of CCSNe. While the high-frequency component might boast a product of distance
and metric strain ($h_+D$) of a few centimeters, the neutrino component
at those low frequencies is $\sim$10-100 centimeters.  However, due to
the much lower frequencies and the squared frequency dependence, the energy
in this component is much smaller ($\sim$a few $\times$ 10$^{-12}$ M$_{\odot}c^2$) than found from the fundamental f-mode
signal due to PNS oscillation \citep{radice2019}.   Nevertheless, the neutrino part
of the CCSNe GW signature at frequencies of 0.1 to 10 Hz could be detected 
by a variety of proposed space- and ground-based GW interferometers \citep{Decihertz,decigo1,decigo2,ET_1,ET_3,aligo}. Stamped on these 
GW data would be information on the asymmetry of matter and neutrino emissions
and characteristic timescales of explosion and neutron star formation 
\citep{braginskii,burrows1996,emuller2012,vartanyan2019}.  Hence,
the low-frequency channels complement those at higher frequencies to 
provide a richer picture of supernova dynamics. This would
complement the science gleaned from the direct neutrino emissions themselves. In fact,
the GW data across the full GW spectral range provides added value $-$ joint 
GW/neutrino analysis would yield returns that are greater than the sum of 
the individual parts.

Simulation studies over the last decade have explored the contribution of matter motions, from convective structure to large-scale instabilities such as the SASI \citep{foglizzo2002,blondin_shaw,foglizzo2012}, to the development of gravitational waves in 2D and 3D simulations and their subsequent detection \citep{vartanyan2019,radice2019,vsg2018,andresen2017,andresen2019,powell2019,powell2020}. Studies have also explored information to be gleaned from detections of neutrino signals with upcoming detectors, such as DUNE and HyperKamiokande \citep{wallace2016,seadrow2018,vartanyan2019,muller2019,kuroda2017}. This progress has paralleled growth in detailed multi-dimensional simulations of CCSNe, many producing robust explosions \citep{vartanyan2018a,vartanyan2018b,radice2017a,burrows_2019,burrows_2019b,nagakura_res,summa2018,oconnor_couch2018b,muller2017,takahashi2019,roberts2016,kuroda2020,ott2018_rel}. 

Less effort, however, has been dedicated to gravitational waveforms from neutrino memory \citep{burrows1996, kotake2007,kotake2009,emuller2012,2013ApJ...766...43M}. Moreover, such studies have either been 2D simulations, or 3D with parameterized neutrino heating. The contribution to gravitational waves from neutrino anisotropies have not been recently studied despite the fact that gravitational waves from CCSNe provide promising candidates for third generation gravitational-wave detectors \citep{Cavagli__2020,Decihertz,srivastava2019,schmitz2020new}.

\begin{figure*}
    \centering
    \includegraphics[width=0.47\textwidth]{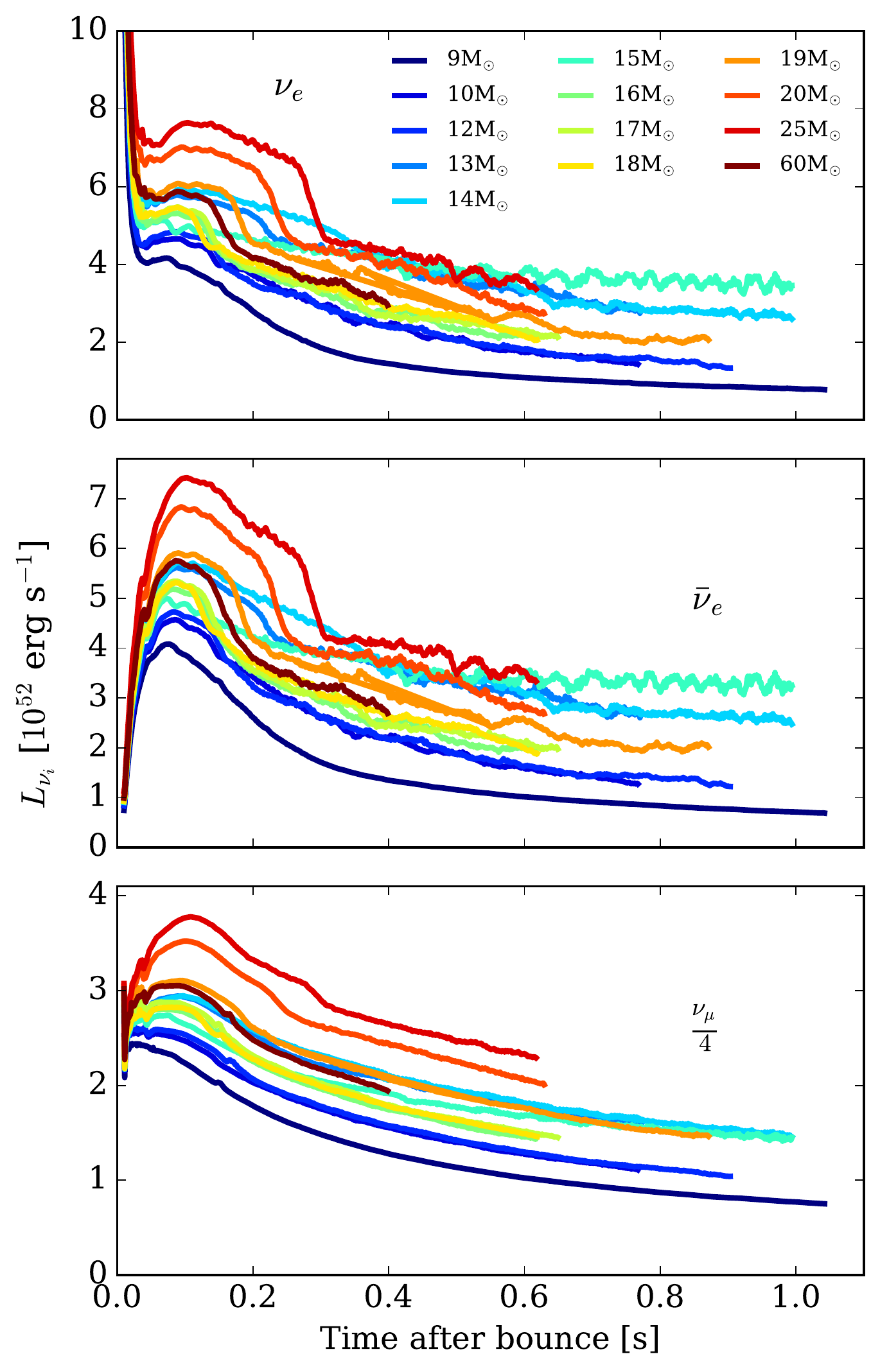}
    \includegraphics[width=0.49\textwidth]{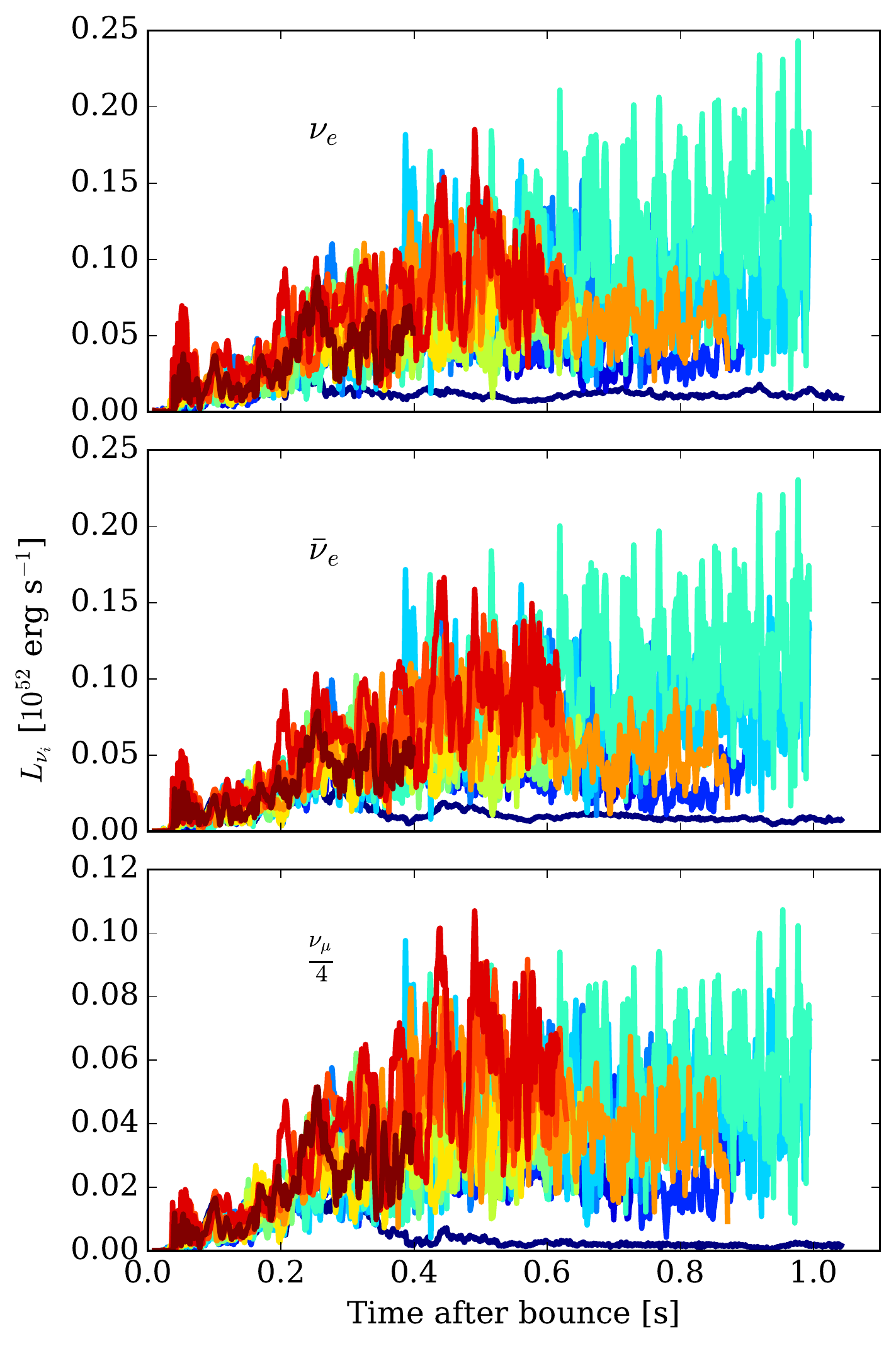}
    \caption{\textbf{Left}: Luminosity (in 10$^{52}$ erg s$^{-1}$) as a function of time after bounce (in seconds)at 250 km and (\textbf{right}): the quadrupole components of the luminosity (in 10$^{52}$ erg s$^{-1}$) as a function of time after bounce (s) for all three neutrino species. The higher luminosities for the electron neutrinos and anti-neutrinos for the 13-, 14-, and 15-M$_{\odot}$ progenitors result from sustained accretion for these non-exploding models. The spiral SASI is most evident for the 15-M$_{\odot}$ progenitor, which shows periodic oscillations in the neutrino luminosities after $\sim$500 ms. The non-exploding models also show large quadrupolar components at later times. However, this does not always yield a higher neutrino gravitational signal. See text for a discussion.}
    \label{fig:lum_quad}
\end{figure*}

Here, we explore gravitational wave signatures from neutrino anisotropies for 13 progenitors from 9$-$60 M$_{\odot}$, all evolved in 3D with our code {F{\sc{ornax}} \citep{skinner2019}. These models were described in \cite{burrows_2019,burrows_2019b}. The paper is outlined as follows: in Sec\,\ref{sec:setup}, we summarize the numerical setup of our simulations and introduce the mathematical framework for studying gravitational waveforms from neutrino anisotropies. In Sec.\,\ref{sec:analysis}, we describe our results 
and explore gravitational waves from neutrino anisotropies as probes of turbulent CCSNe dynamics. 
We discuss the frequency dependence of the GW signal and the resulting prospects for future detection.
We present our conclusions in Sec\,\ref{sec:conc}.

\section{Numerical Setup}\label{sec:setup}

F{\sc{ornax}} \citep{skinner2019} is a multi-dimensional, multi-group radiation hydrodynamics code originally constructed to study core-collapse supernovae. It features an M1 solver for neutrino transport with detailed neutrino microphysics and a monopole approximation to general relativity (\citealt{marek_janka2009}). In this paper, we study 13 stellar progenitors in 3D, with ZAMS (Zero Age Main Sequence) masses of 9, 10, 12, 13, 14, 15, 16, 17, 18, 19, 20, 25, and 60 M$_{\odot}$ models. All models are initially collapsed in 1D through 10 ms after bounce, and 
then mapped to three dimensions. For all progenitors except the 25-M$_{\odot}$ models, we use \cite{swbj16}. The 25-M$_{\odot}$ progenitor is from \cite{sukhbold2018}. The models and setup here are identical to those discussed in \citep{burrows_2019b,vartanyan2019}. We note that all models except the 13-,14-, and 15-M$_{\odot}$ progenitors explode \citep{burrows_2019b}. Additionally, we note the F{\sc{ornax}} treats three species of neutrinos: electron-neutrinos ($\nu_e$), electron anti-neutrinos ($\bar{\nu}_e$), and $\mu$,$\tau$ neutrinos and their antiparticles bundled together as heavy-neutrinos, ``$\nu_\mu$".

To calculate gravitational waves from neutrino asymmetries, we follow the prescription of \cite{emuller2012}. We include angle-dependence of the observer through the viewing angles $\alpha \in [-\pi,\pi]$ and $\beta \in [0,\pi]$. The time-dependent neutrino emission anisotropy parameter for each polarization is defined as 

\begin{equation}\label{eq:0}
\alpha_S (t,\alpha,\beta) = \frac{1}{\Lambda(t)} \int_{4\pi} d\Omega'\, W_S(\Omega',\alpha,\beta) \frac{d\Lambda}{d\Omega'}(\Omega',t)\,,
\end{equation} where the subscript $S \in \{+,\times\}$ and the gravitational wave strain from neutrinos is defined as 

\begin{equation} \label{eq:1}
h_S(t,\alpha,\beta) = \frac{2G}{c^4D} \int_0^t dt'\, \Lambda(t')\,\alpha_S(t',\alpha,\beta)\,,
\end{equation} where $\Lambda (t)$ is the angle-integrated neutrino luminosity as a function of time, $D$ is the distance to the source, $\Omega'$ is the angular differential in the coordinate frame of the source, and W$_\mathrm{S}$($\mathrm{\Omega}',\alpha,\beta)$ is the geometric weight for the anisotropy parameter given by  

\begin{equation}
    W_{\mathrm{S}} = \frac{D_{\mathrm{S}} (\theta',\phi',\alpha,\beta)}{N(\theta',\phi',\alpha,\beta)}\,,
\end{equation} where 

\begin{subequations}
\begin{align}
\begin{split}
D_+ &= [1 + (\cos(\phi')\cos(\alpha) + \sin(\phi')\sin(\alpha))\sin(\theta')\sin(\beta)\\&+\cos(\theta')\cos(\beta)]\{[(\cos(\phi')\cos(\alpha)+\sin(\phi')\sin(\alpha))\\&\sin(\theta')\cos(\beta)-\cos(\theta')\sin(\beta)]^2-\sin^2(\theta')(\sin(\phi')\cos(\alpha)\\&-\cos(\phi')\sin(\alpha))^2\}
\end{split}\\
\begin{split}
D_\times &= [1 + (\cos(\phi')\cos(\alpha) + \sin(\phi')\sin(\alpha))\sin(\theta')\sin(\beta)\\&+\cos(\theta')\cos(\beta)]\,\,2\,[(\cos(\phi')\cos(\alpha)+\sin(\phi')\sin(\alpha))\\&\sin(\theta')\cos(\beta)-\cos(\theta')\sin(\beta)]\sin(\theta')(\sin(\phi')\cos(\alpha)\\&-\cos(\phi')\sin(\alpha))^2
\end{split}\\
\begin{split}
N &= [(\cos(\phi')\cos(\alpha)+\sin(\phi')\sin(\alpha))\sin(\theta')\cos(\beta)\\&-\cos(\theta')\sin(\beta)]^2+\sin^2(\theta')(\sin(\phi')\cos(\alpha)\\&-\cos(\phi')\sin(\alpha))^2\,.
\end{split}
\end{align}
\end{subequations}

\section{Results and Discussion} \label{sec:analysis}

In Fig.\,\ref{fig:lum_quad}, we plot on the left-hand side the neutrino luminosity (equivalent to $\Lambda(t)$ for each neutrino species in Eqns\,\ref{eq:0}-\ref{eq:1}) 
as a function of time for all 3D models considered here for all three neutrino species. 
Lower mass models, such as the 9-M$_{\odot}$ model in particular, typically have lower 
neutrino luminosities and accretion rates.  Absent explosion, the 13-,14-, and 15-M$_{\odot}$ 
models accrete for longer and evince higher neutrino luminosities at later times. 
The development of the spiral SASI \citep{blondin2003,blondin_shaw,blondin2005,kuroda2016} 
is evident for these non-exploding models through the quasi-periodic oscillations 
after $\sim$500 ms in their neutrino luminosities, and is most visible for the electron neutrinos ($\nu_e$) 
and anti-neutrinos ($\bar{\nu}_e$).  We note that we see only the spiral SASI in the 
non-exploding models. The smaller shock stagnation radii inherently manifest in non-exploding models 
favor faster advective-acoustic timescales (\citealt{foglizzo2002}) that efficiently 
amplify SASI growth (see \citealt{scheck2008}). Recent 3D simulations confirm this 
for non-exploding, or delayed-exploding, models 
(\citealt{roberts2016,kuroda2016,kuroda2017,ott2018_rel,glas2018,vartanyan2018b,burrows_2019b,vartanyan2019}). As indicated in \citealt{burrows2018,vartanyan2018a,vartanyan2018b,burrows_2019b}, the majority of our models explode within the first 300 milliseconds, precluding the development of a SASI. Prompt explosion is in large part due to the detailed microphysics of the simulation $-$ namely the inclusion of the axial-vector many-body correction to the neutrino-nucleon scattering rates $-$ and to the intrinsic density profile of the progenitors. Generally, we find that models with sharper Silicon/Oxygen interfaces produce earlier explosion owing to a favorable combination of a sustained neutrino luminosity and a drop in the accreta ram pressure. We also emphasize that the nature of explosion in these models is not absolute $-$ some models seem more explodable than others, but changes to the evolution physics, progenitor structure, or resolution could alter this conclusion.}

In the right hand side of Fig.\,\ref{fig:lum_quad}, we plot the quadrupolar neutrino luminosity as a function of time for all 3D models and for all three neutrino species. We use the approach outlined in \cite{burrows2012} (see also \citealt{vartanyan2019}) to
decompose the luminosity L$_\nu(\theta,\phi)$ into spherical harmonic components with coefficients:
\begin{equation}\label{eq:alm}
a_{lm} = \frac{(-1)^{|m|}}{\sqrt{4\pi(2l+1)}} \oint L_\nu(\theta,\phi) Y_l^m(\theta,\phi) d\Omega\, ,
\end{equation} normalized such that $a_{00} = a_{0} =\langle L_\nu\rangle$ (the average
neutrino luminosity). $a_{11}$, $a_{1{-1}}$, and $a_{10}$ correspond to the average
Cartesian coordinates of the shock surface dipole $\langle x_s\rangle$, $\langle y_s\rangle$,
and $\langle z_s\rangle$, respectively.
The orthonormal harmonic basis functions are given by
\begin{equation}
Y_l^m(\theta,\phi) = \begin{cases}
        \sqrt{2} N_l^m P_l^m(\cos\theta) \cos m\phi&            m>0\, ,\\
        N_l^0 P_l^0(\cos\theta) &                               m=0\, ,\\
        \sqrt{2} N_l^{|m|} P_l^{|m|}(\cos\theta) \sin |m|\phi&  m<0\, ,
\end{cases}
\end{equation}
where
\begin{equation}
N_l^m = \sqrt{\frac{2l+1}{4\pi}\frac{(l-m)!}{(l+m)!}}\, ,
\end{equation} $P_l^m(\cos\theta)$ are the associated Legendre polynomials,
and $\theta$ and $\phi$ are the spherical coordinate angles.

We define the quadrupolar component of the neutrino luminosity as the following norm,

\begin{equation}
L_{\nu,\ell} = \frac{\sqrt{\sum_{m=-\ell}^{\ell} a_{\ell m}^2}}{a_{00}}\,,
\end{equation} where we take $\ell=2$.

The quadrupolar neutrino luminosity is at most a few percent of the total neutrino luminosity. Note that the exploding models peak at $\sim$0.5 second after bounce, after which the non-exploding models develop the largest quadrupolar components due to the onset of the spiral SASI. The weakly-exploding 9-M${\odot}$ progenitor shows the smallest asymmetry.

\begin{figure*}
    \centering
    \includegraphics[width=0.49\textwidth]{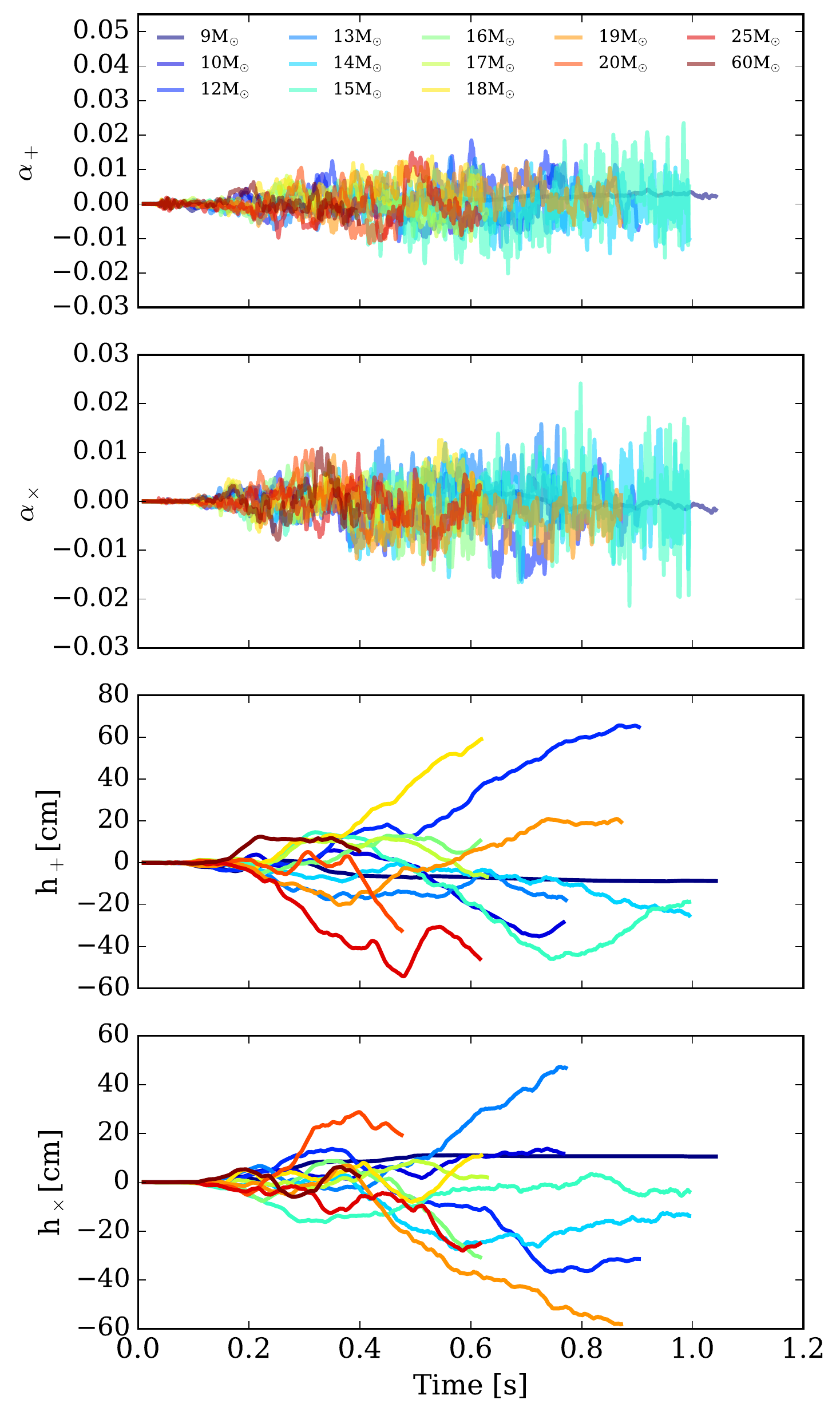}
    \caption{We show the neutrino anisotropy parameters for the electron-type neutrinos for both polarizations (top two panels), and the gravitational wave strain summed over all species of neutrinos for both polarizations (bottom panels). The values are measured along the negative x-axis.}
    \label{fig:strain_nu}
\end{figure*}

In Fig.\,\ref{fig:strain_nu}, we show the geometric anisotropy parameter along the negative x-axis for the electron-type neutrinos for the $+$ polarization (top panel) and the $\times$ polarization (second panel from the top). The neutrino anisotropies can get as strong as several percent, in accordance with the quadropolar component of the neutrino luminosity. This is a factor of several higher than for the 3D simulations from \cite{emuller2012}. Although the ``heavy"-neutrinos show similar or slightly smaller geometric anisotropy parameters, they dominate the neutrino luminosity and hence the gravitational wave strain, plotted for $+$ and $\times$ polarizations in the bottom two panels, respectively, summed over all neutrino species. Note that we see a similar hierarchy between neutrino species and their spatial variation here as in \cite{vartanyan2019}, using different formalisms for the neutrino anisotropy. The anisotropy parameters for the non-exploding models are largest for all species at late times, after $\sim$500 ms, corresponding to the development of the spiral SASI. However, a larger anisotropy parameter doesn't directly translate into higher strains, as visible in the bottom two panels of Fig.\,\ref{fig:strain_nu}, due to additive cancellation when integrating Eq.\,\ref{eq:1} over time. 

The neutrino gravitational strain is approximately two orders of magnitude larger than the matter gravitational wave strain and shows secular growth with time. The neutrino strain is significantly different from the matter component \citep{radice2019,vsg2018} $-$ it features weaker time variations and larger amplitudes, with more monotonic behavior with time. This reflects the fundamental differences of their origins. Gravitational waves from matter involve $\sim$0.1 M$_{\odot}$ with convective velocities of $\sim$1000 km s$^{-1}$ on convective timescales of milliseconds \citep{vartanyan2018b,nagakura_pns}. On the other hand, neutrino luminosity contributions involve relativistic velocities and significant energy losses from the PNS. The memory effect \citep{braginskii} from neutrinos, indicated as the integral over time in Eq.\,\ref{eq:1}, smooths over small time-scale variations and allows cumulative growth of the neutrino strain. 

We comment on the 9-M${\odot}$ progenitor, which explodes early and more isotropically than later exploding models, such as the 19- and 60-M${\odot}$ progenitors \citep{burrows_2019b}. Early, spherical explosion and the cessation of accretion was associated with a weaker gravitational wave signal from matter motions.  \cite{radice2019} emphasized that accretion, and not solely PNS convection, was responsible for driving matter gravitational waves. The 9-M$_{\odot}$ is a good example of this distinction $-$ the PNS is convective at late times \citep{nagakura_pns}, but accretion has ended. Here, we further associate the cessation of accretion with a weak GW signal from neutrino anisotropies. The 9-M$_{\odot}$ progenitor has both small anisotropy parameters in Fig.\,\ref{fig:strain_nu} and small strains.

Note that all models take $\sim$100-200 ms for the neutrino anisotropy to manifest and result in a non-zero strain. This corresponds to the timescale for turbulence to develop and for the supernova shock to break spherical symmetry. Note that, unlike the matter component of gravitational waves, neutrinos do not show a prompt-convection burst shortly after bounce. However, like the matter component, there is a hiatus of $\sim$100-200 ms. 

Lastly, we see large excursions from monotonic growth of the strain for several models due to the delay until fully-developed turbulence arises. Neglecting azimuthal-variations for the sake of simplicity, we can understand the strain evolution through the evolving geometry of the explosion. We take the $+$ polarization for an observer along the negative x-axis, where positive strains correspond to axial motions, and negative strains to equatorial motions (see also \citealt{kotake2009} for a 2D analogue along the equatorial plane). The 25-M${\odot}$ progenitor has a large negative strain until $\sim$400 ms post-bounce, when the model explodes. We see a corresponding positive bump in the $\alpha_+$ in the top panel of Fig.\,\ref{fig:strain_nu}, and a positive bump in h$_+$ which corresponds to the deformation of the shock as it begins to expand. Additionally, the non-exploding progenitors maintain small strains until $\sim$500 ms post-bounce, when the spiral SASI develops. The 15-M$_{\odot}$ progenitor (in cyan) develops a large negative h$_+$ thenabouts, corresponding to an equatorial spiral SASI. Within a few hundred milliseconds, the strain becomes less negative as the spiral SASI precesses. We emphasize that this conclusion depends on the observer direction.

\begin{figure*}
    \centering
    \includegraphics[width=0.9\textwidth]{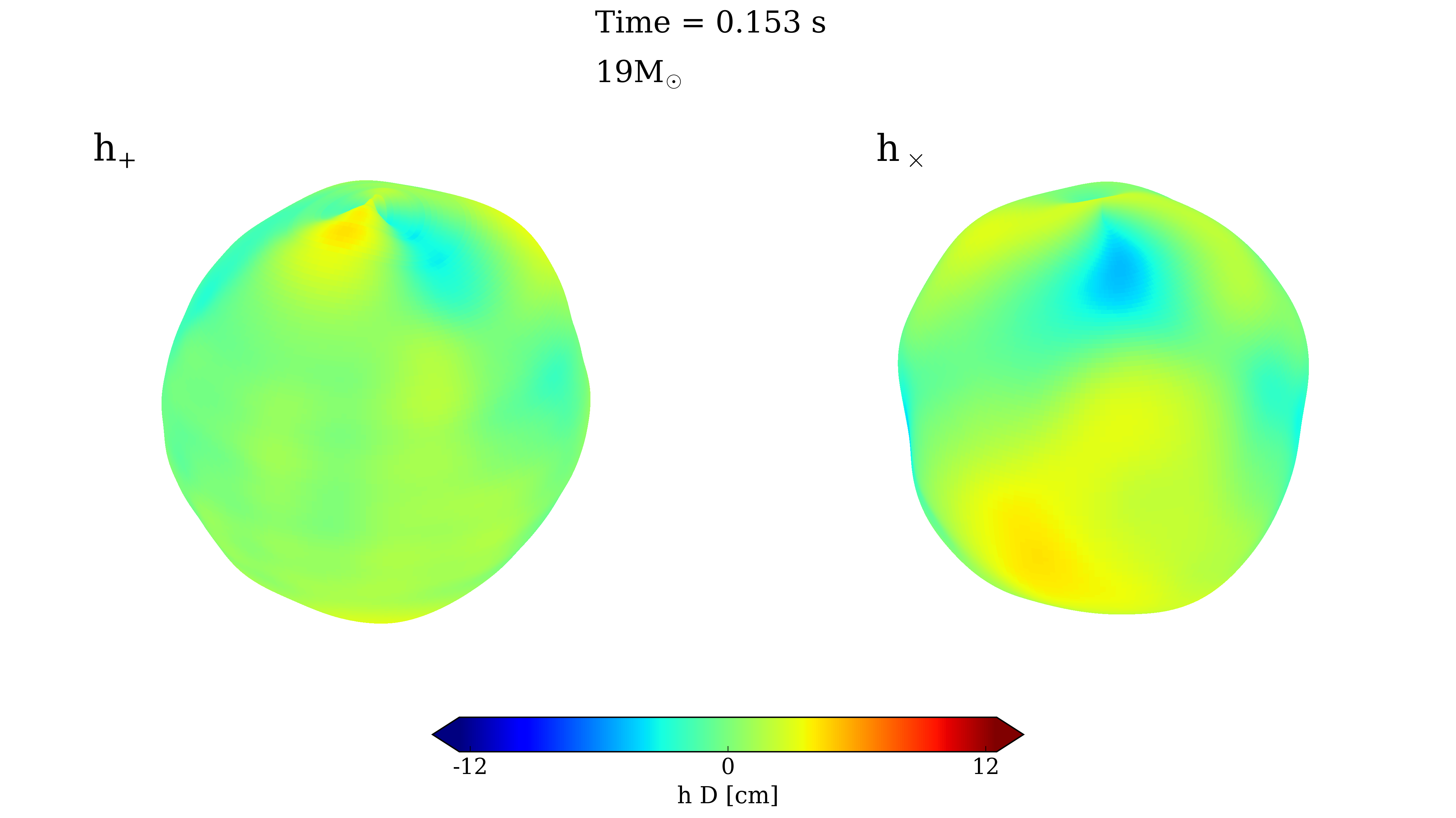}
    \includegraphics[width=0.9\textwidth]{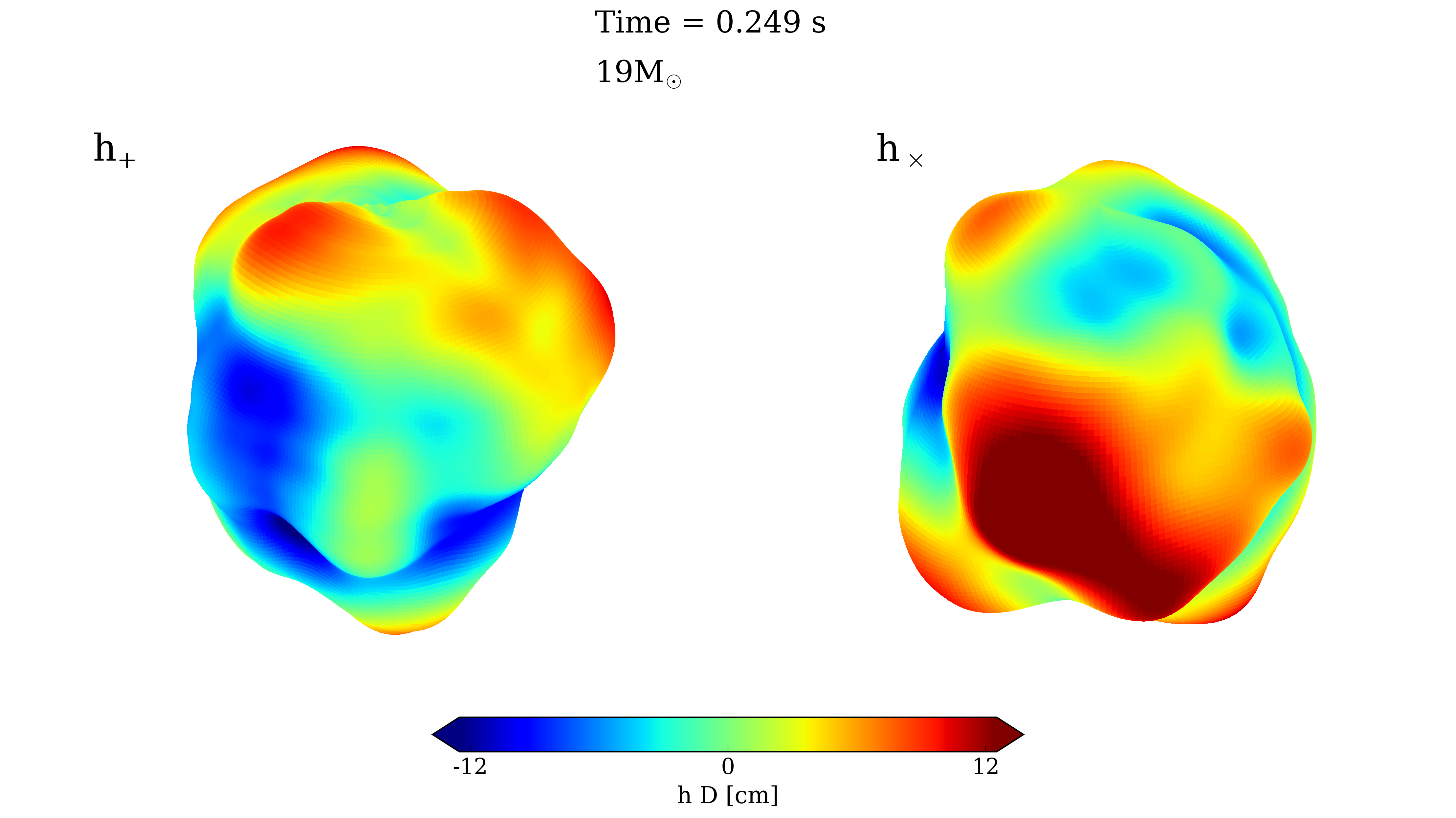}
    \caption{Three-dimensional map illustrating the gravitational-wave strain $h$ (multiplied by distance $D$) generated by neutrino emission anisotropies $253~\mathrm{ms}$ after bounce, assuming a stellar progenitor with mass 19-M$_{\odot}$. The signal is shown for both $h_+$ (left) and $h_\times$ (right) polarization, and as a function of the viewing angle. Hotter colors (yellow to red; convex surfaces), indicate positive strains, whereas cooler colors (blue to yellow; concave surfaces) indicate negative strains.}
\label{fig:GW_map}
\end{figure*}

\begin{figure*}
    \centering
    \includegraphics[width=0.9\textwidth]{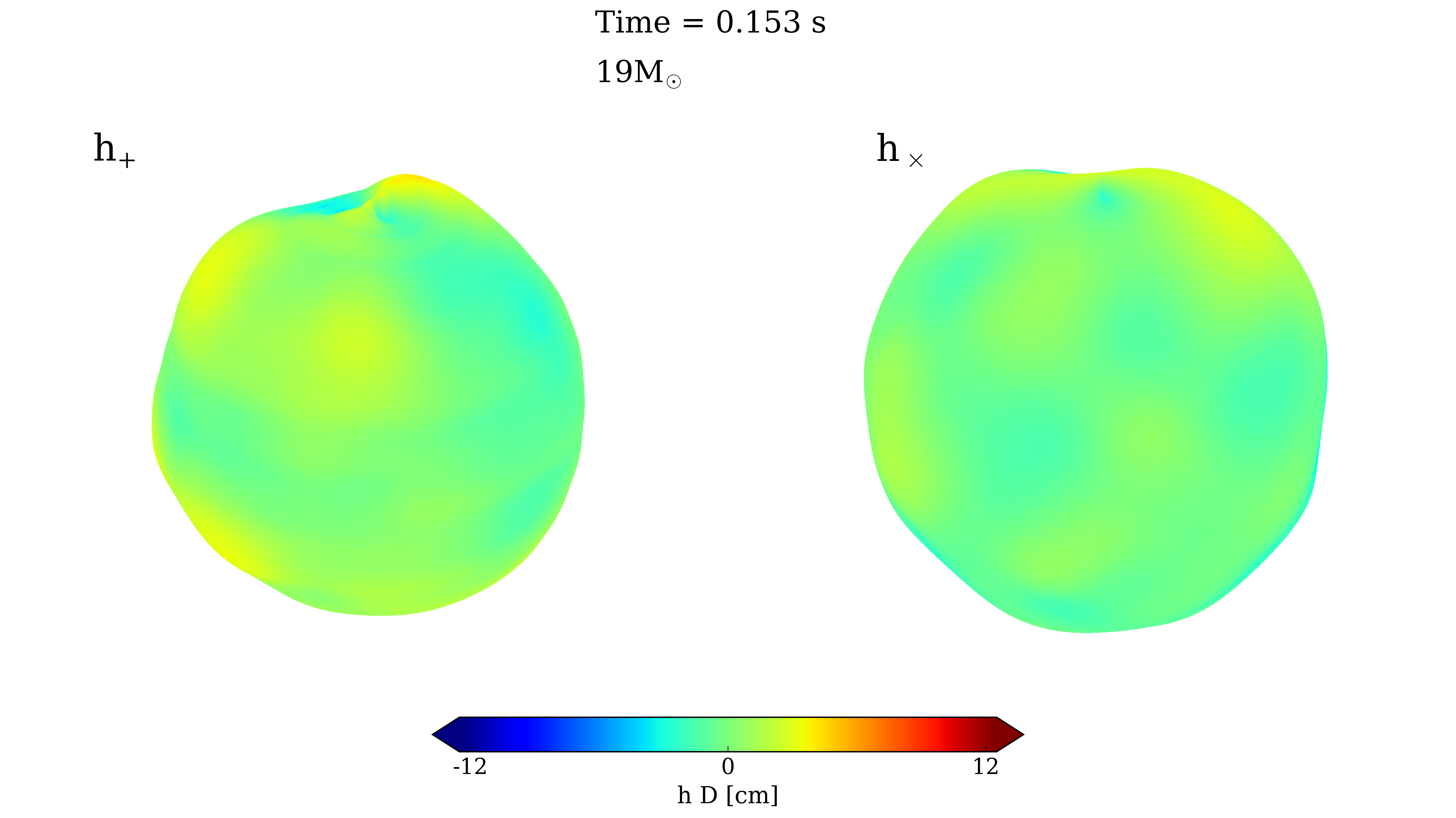}
    \includegraphics[width=0.9\textwidth]{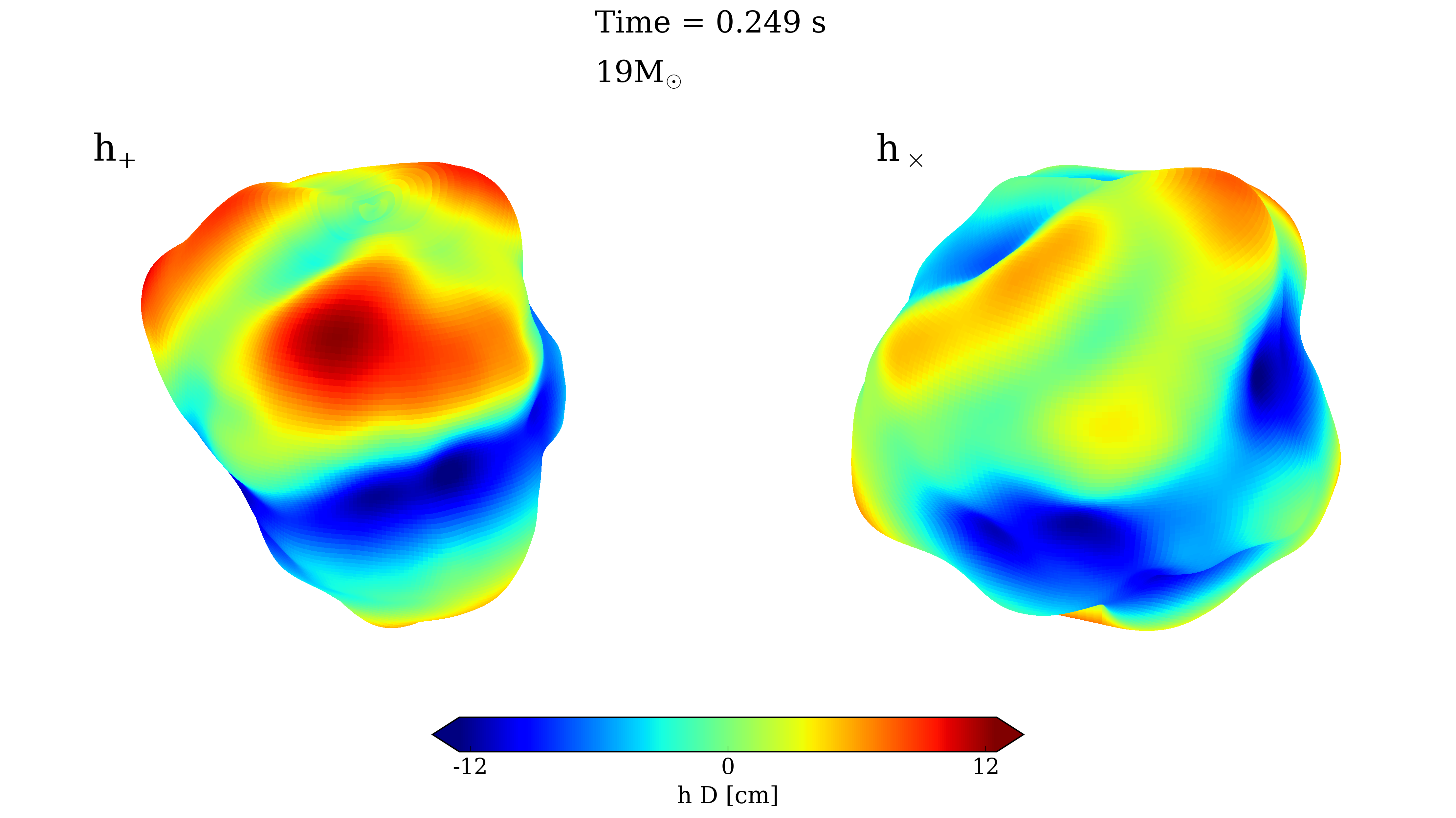}
    \caption{Same as Fig.\,\ref{fig:GW_map}, but now with the viewing angle rotated by 180$^{\circ}$ along the azimuth.}
\label{fig:GW_map_2}
\end{figure*}

We illustrate the three-dimensional gravitational wave emission from neutrinos for both 
polarizations in Fig.\,\ref{fig:GW_map} at 153 ms and 249 ms post-bounce for the 19-M$_{\odot}$ 
progenitor as a function of viewing angle. Here, colors are degenerate with the contours: hotter 
colors and convex surfaces indicate larger positive values of the strain, and cooler colors 
and concave surfaces indicate larger negative values of the strain. To emphasize the dependence 
on viewing-angle, we illustrate in Fig.\,\ref{fig:GW_map_2} the same map of the gravitational 
wave-emission rotated by 180$^{\circ}$ in azimuth. The prompt rise in the gravitational wave energy from matter contributions at $\sim$50 ms corresponds to prompt convection following neutrino breakout. The energy growth then stalls for $\sim$100 ms, before reaching values as high as $\sim$10$^{46}$ erg, and less than $\sim$10$^{44}$ erg for the lowest energy, for the 9-M$_{\odot}$ model. The stall time for the gravitational wave energy to ramp up indicates the timescale for turbulence to develop in the core-collapse supernova. This timescale for matter is not dissimilar for that from neutrinos: $\sim$100 ms. On the other hand, the energy in gravitational waves from neutrinos can be as high as 10$^{43}$ erg, and as low as 10$^{41}$ erg, again for the 9-$M_{\odot}$ model. We note that these values are still increasing at the end of our simulation, as is the explosion energy, and longer 3D simulations are required to capture the asymptotic behavior. Additionally, non-exploding models have powerful gravitational wave signatures in both neutrinos and matter at late times that is associated with the development of the spiral SASI, which modulates the infalling accretion to source gravitational waves.

\begin{figure*}
    \centering
    \includegraphics[width=0.49\textwidth]{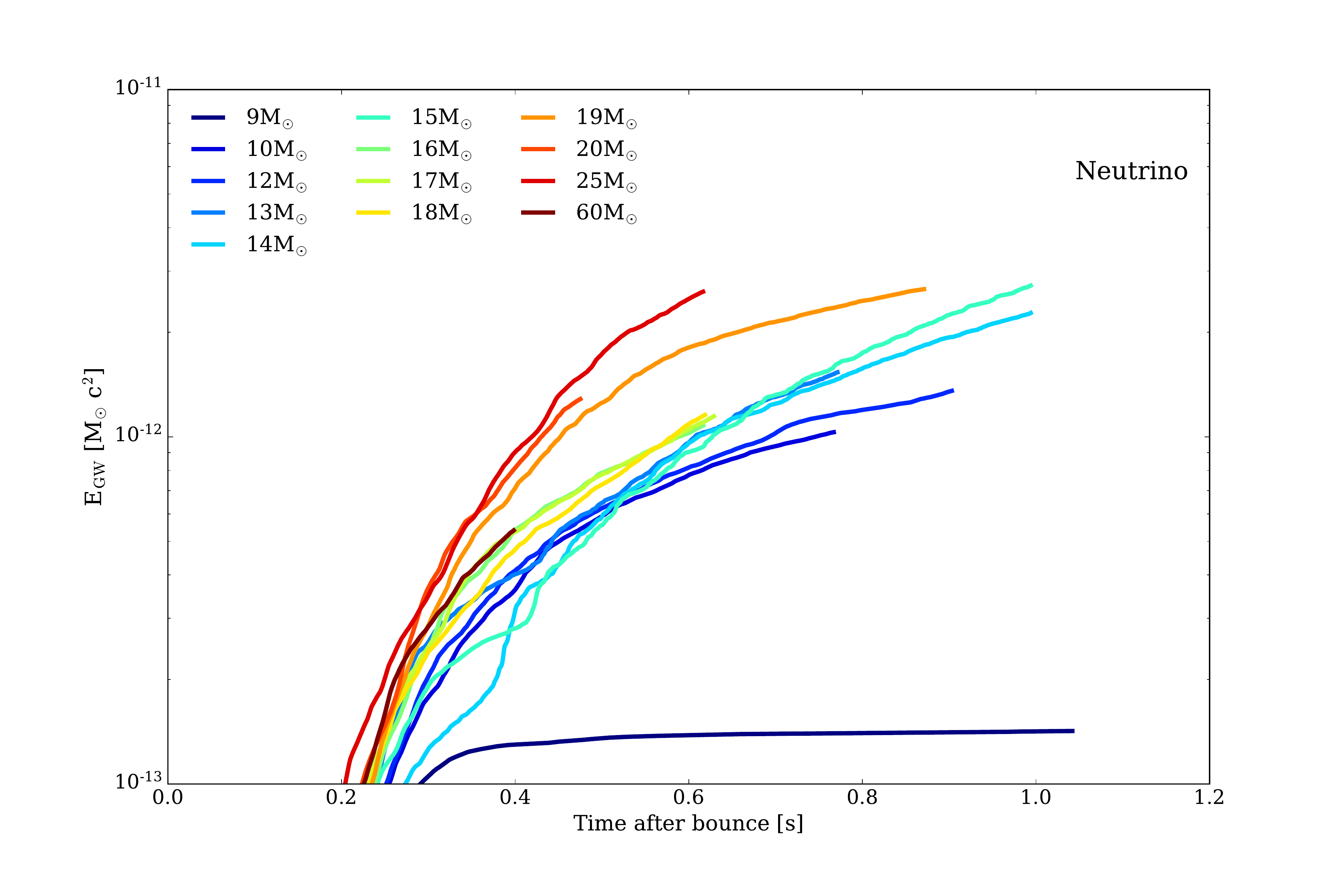}
    \hfill   
    \includegraphics[width=0.49\textwidth]{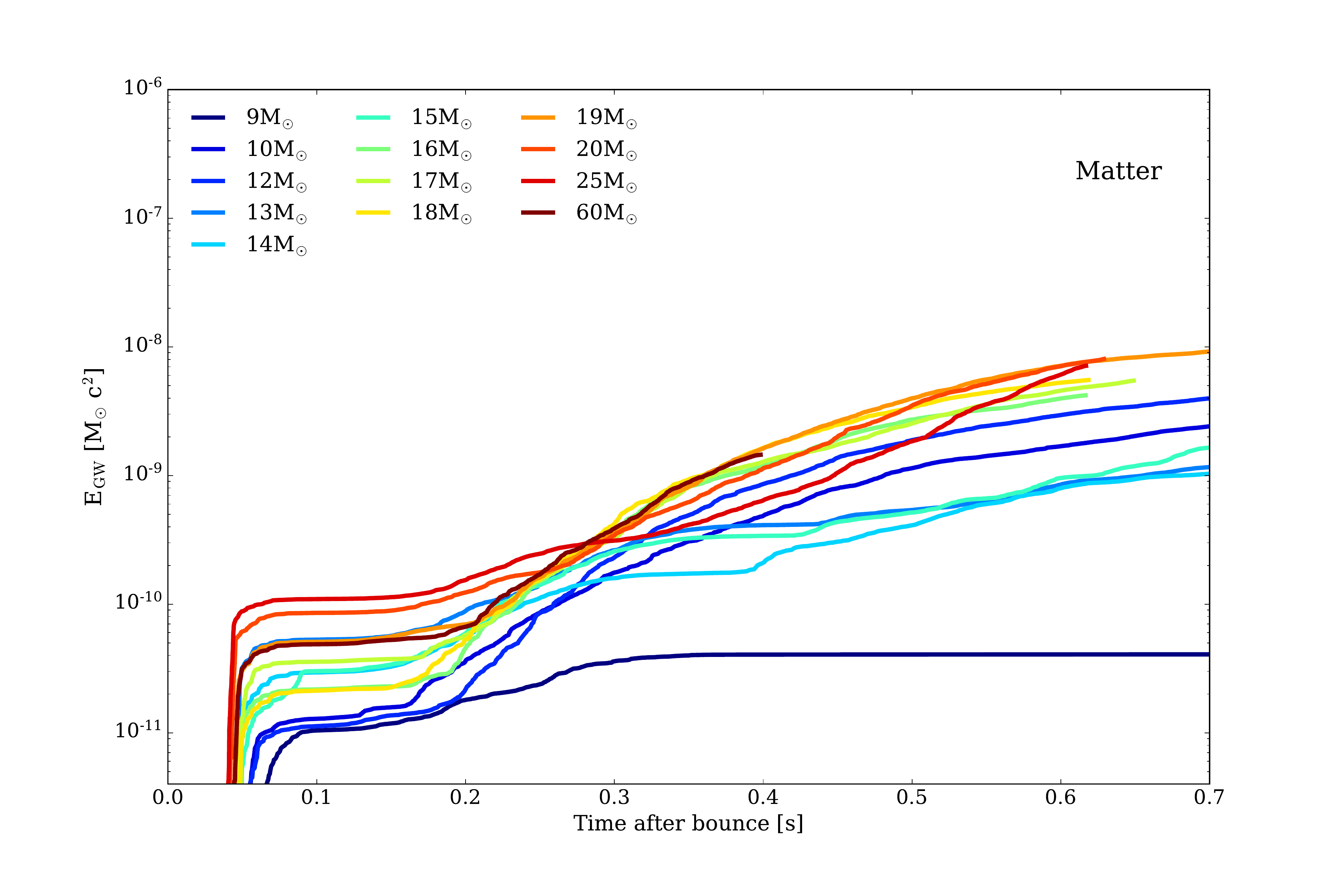}
    \caption{Here, we compare the contribution to the gravitational wave energy (in M$_{\odot}$c$^2$) from neutrino (left) and matter (right) quadrupolar asymmetries for the various models studied in 3D as a function of time after bounce (in seconds). Note the vastly different scales on the y-axis. Neutrino contributions to the gravitational wave energies are more than three orders of magnitude smaller than those due to
    matter motions.}
    \label{fig:EGW}
\end{figure*}

In Fig.\,\ref{fig:EGW}, we show the gravitational wave energies from neutrino anisotropies on the left for all our models, and from matter on the right. Note that, although neutrinos dominate the gravitational strain by as much as two orders of magnitude, matter dominates the gravitational wave energy by as much as three orders of magnitude. This is simply because gravitational wave energy scales as the square of the product of the strain and the frequency, and the neutrino component resides at much lower frequencies (see Fig.\,\ref{fig:spec}), indicative of its secular evolution and the memory effect.

\subsection{Frequency Dependence}
\begin{figure*}
    \centering
    \includegraphics[width=0.9\textwidth]{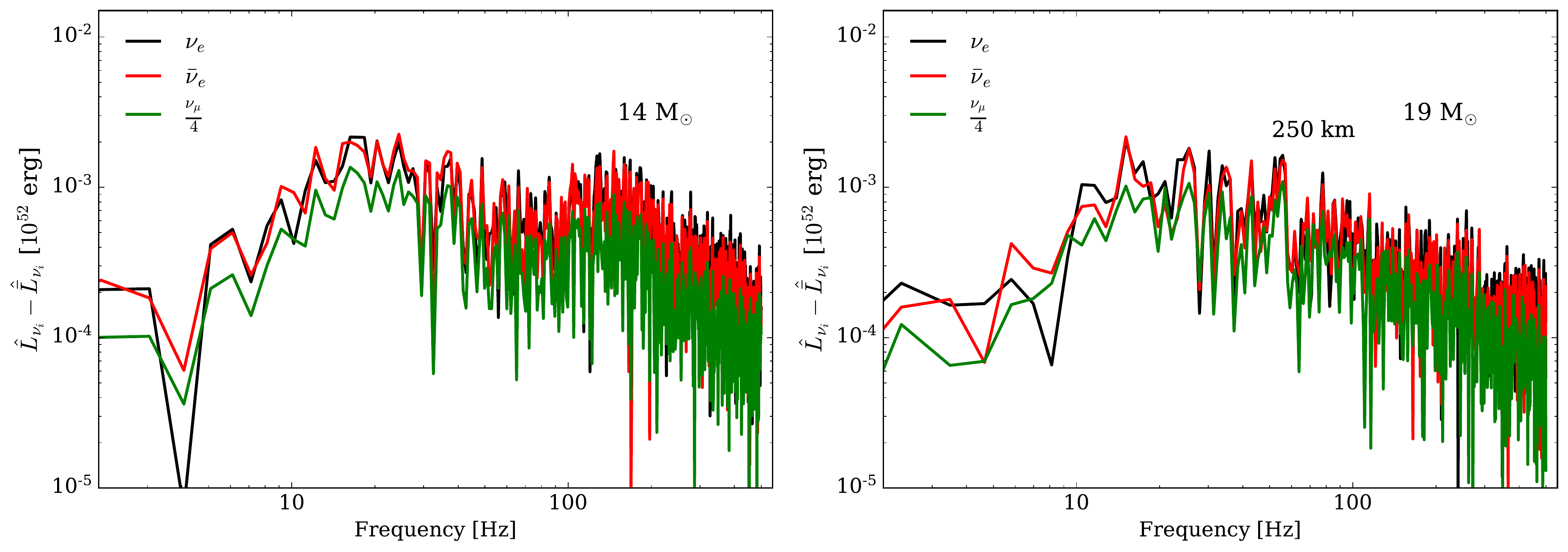}
\caption{Fourier transform of the neutrino luminosity for all three species, subtracting out the mean over a 30-ms running average, as a function of frequency (Hz). We see significant power at low frequencies corresponding the longer secular timescales of order one second. The 14-M$_{\odot}$ progenitor (left), which does not explode, shows a significant bump in power at $\sim$150 Hz, corresponding to the development of the spiral SASI. The 19-M$_{\odot}$ progenitor (right), which does explode, is absent the SASI and the corresponding neutrino signature. Both models show also a slight bump at $\sim$300 Hz, corresponding to the timescales of small-scale convection in the PNS core.}
    \label{fig:quad_FFT}
\end{figure*}

In Fig.\,\ref{fig:quad_FFT}, we provide Fourier transforms of the neutrino luminosity for 
all three species, subtracting out the mean over a 30-ms running average, as a function 
of frequency (in Hz) for the 14-M$_{\odot}$ progenitor (left) and the 19-M$_{\odot}$ 
progenitor (right). All neutrino species show similar frequency behavior with a weak 
hierarchy in power in the order $\nu_e$, $\bar{\nu}_e$, and $\nu_{\mu}$, similar to 
the results in \protect{\cite{vartanyan2019}}. We see significant power at low frequencies corresponding to the longer secular timescales of order one second. The 14-M$_{\odot}$ progenitor, which does not explode, shows a significant bump in power at $\sim$150 Hz, corresponding to the development of the spiral SASI. Such a low-power component, between 100-200 Hz, has been well-identified with the SASI, which has modulation timescales of $\sim$O(10 ms) \citep{blondin_shaw,2009ApJ...697L.133K,kuroda2016,kuroda2017,andresen2017}. The exploding 19-M$_{\odot}$ progenitor has no SASI development.

\begin{figure*}
    \centering
    \includegraphics[width=0.5\textwidth]{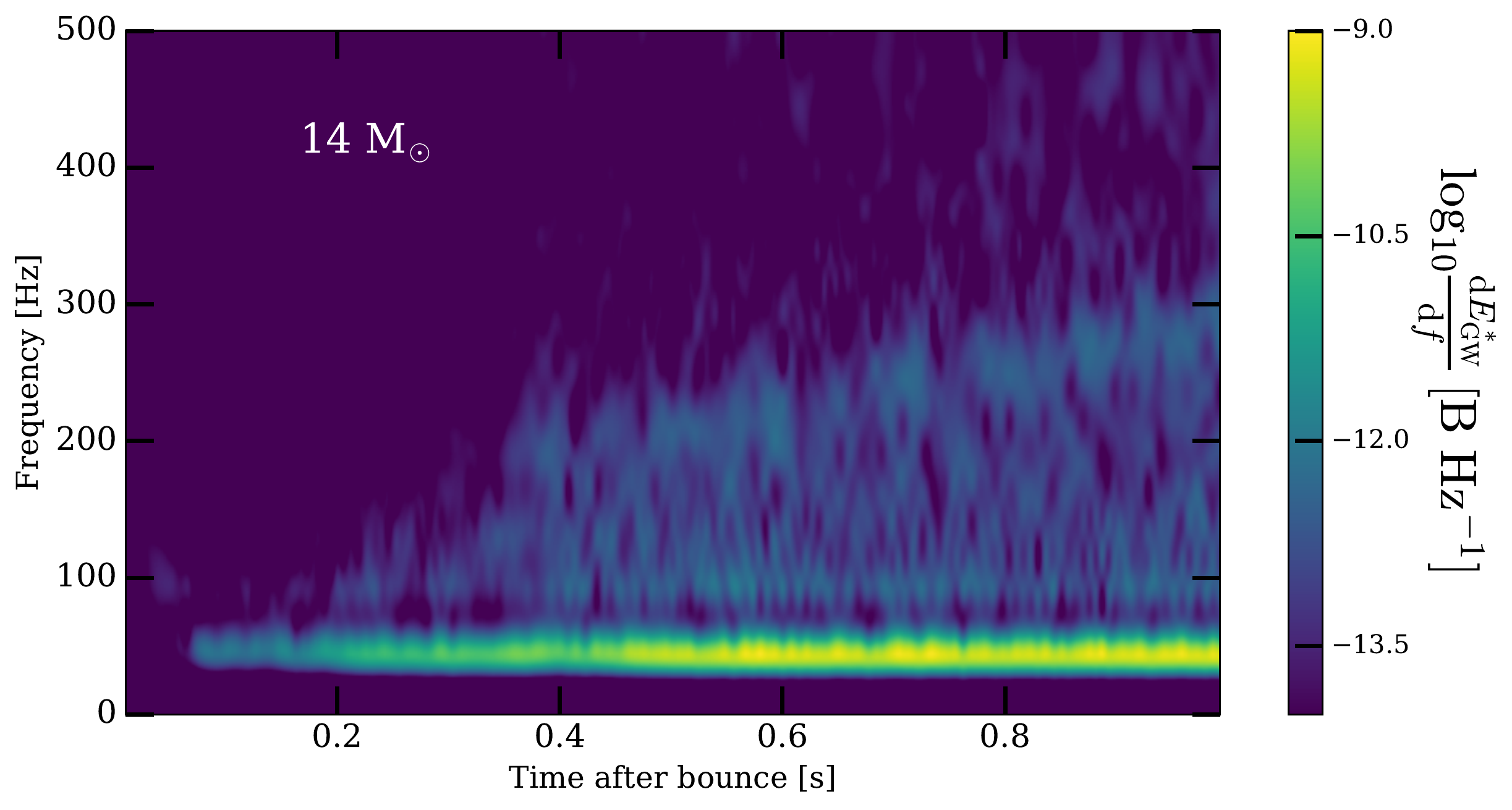}\hfill
    \includegraphics[width=0.5\textwidth]{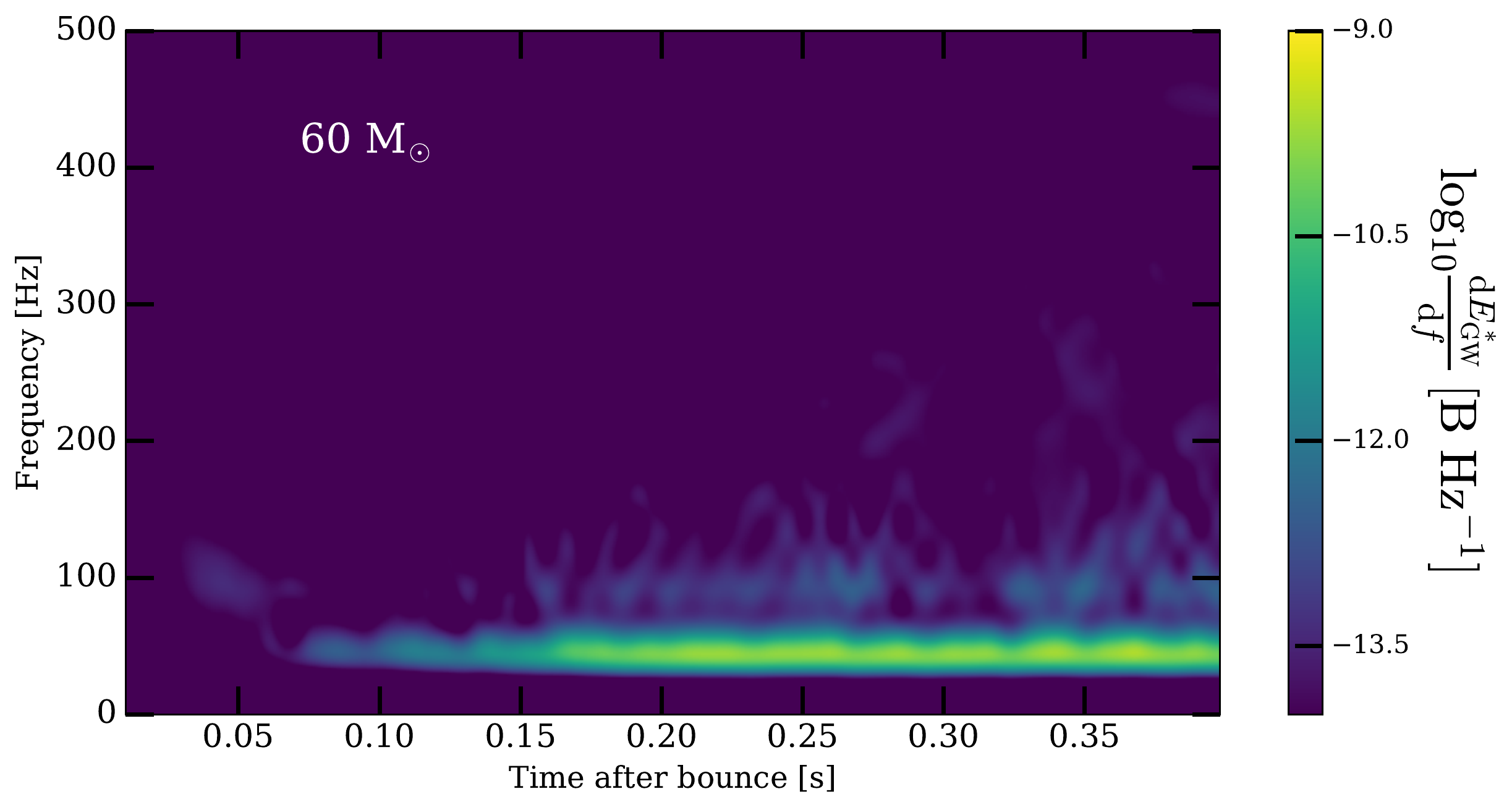}    
    \caption{Gravitational wave energy spectrogram (in B Hz$^{-1}$) from neutrino anisotropies of the non-exploding 14-M$_{\odot}$ progenitor (left), and the exploding 60-M$_{\odot}$ progenitor (right) as a function of time after bounce (s) and frequency (Hz). Note that most power lies below 50 Hz, with less power at higher frequencies. The spiral SASI in non-exploding models, like the 14-M$_{\odot}$ progenitor, contributes to this power.}
    \label{fig:spec}
\end{figure*}

In Fig.\,\ref{fig:spec}, we provide gravitational wave energy spectrograms (in B Hz$^{-1}$) from neutrino anisotropies of the non-exploding 14-M$_{\odot}$ progenitor (left), and the exploding 60-M$_{\odot}$ progenitor (right) as a function of time after bounce (s) and frequency (Hz). To compute the spectograms, we apply a Hann window function to the gravitational wave 
energy spectra and take a short time Fourier transform with a window size of 40 ms. Our 
sampling frequency is 1000 Hz. Note that most of the power lies below 50 Hz, with less 
power at higher frequencies. The spiral SASI in non-exploding models, like the 14-M$_{\odot}$ 
progenitor, contributes to this higher frequency power.

\subsection{Detection Prospects}
\begin{figure*}
    \centering
    \includegraphics[width=0.88\textwidth]{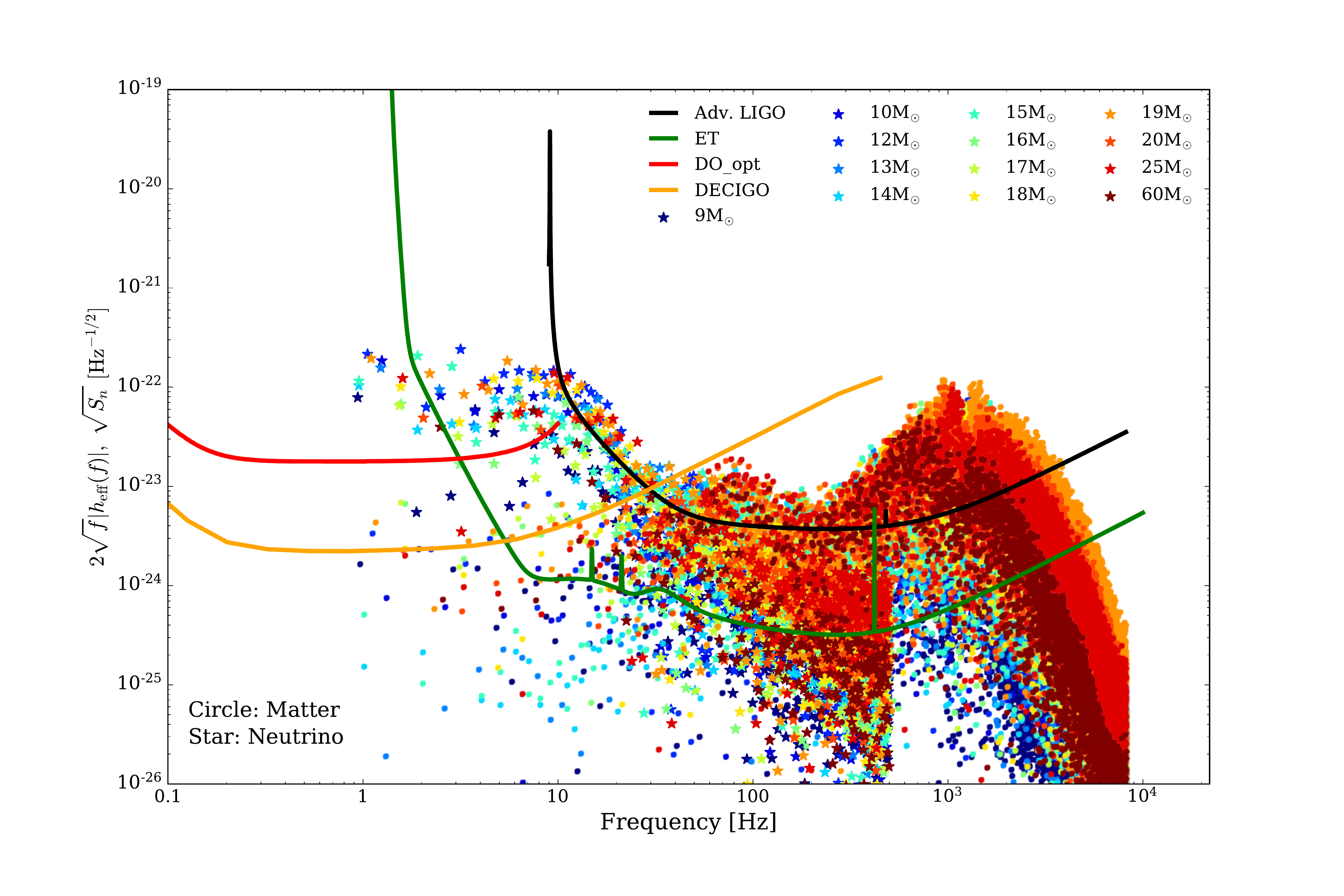}
    \caption{We plot the amplitude spectral density at 10 kpc (in Hz$^{-1/2}$) spanning $\sim$1$-$10,000 Hz for all the models studied in 3D. The neutrino component (stars) dominates from sub-Hz to several hundred Hz, whereas the matter component (circles) dominates above several hundred Hz. We also overplot the sensitivity curves for current and upcoming gravitational wave detectors. Many detectors, including aLIGO, DECIGO, and ET, will be able to detect gravitational waves from a galactic CCSN event. DO$\_$opt indicates an optimal decihertz obervatory between 2035-2050 \protect\citep{Decihertz} lying between LISA and ground-based detectors in frequency band designed to detect intermediate mass black hole binaries, a few $\sim10$s to a few $\sim$100s M$_{\odot}$, with the added value of detection capabilities for galactic CCSNe.}
    \label{fig:sens_3D}
\end{figure*}

In Fig.\,\ref{fig:sens_3D}, we plot the gravitational wave amplitude spectral density at 10 kiloparsecs for all models studied in 3D, indicating both the neutrino component (stars) and matter component (circles) and compare with the sensitivity of current and proposed GW missions. Although we show the matter and neutrino components separately, in reality they combine into the net gravitational wave strain. However, as illustrated, they dominate at different frequency ranges. Above $\sim$100 Hz, the matter component dominates, peaking at 1000 Hz, corresponding to convective timescales. The neutrino component dominates below 100 Hz and plateaus below a few tens of hertz. 

 We overplot the sensitivity curves for three detectors: Advanced-LIGO (Zero-Det High-P design), the Einstein Telescope (design D), and DECIGO.
Advanced-LIGO will be able to resolve GWs for a galactic CCSN from $\sim$ten to a few thousand Hertz \citep{aligo}. The Einstein Telescope \citep{ET_1,ET_3}, a ground-based detector with a triangle distribution of arms with baselines of ten kilometers (as opposed to four for aLIGO) will provide improved sensitivity down to one Hertz, capable of detecting even lower mass progenitors with weaker GW signals. DECIGO is a space-based heliocentric mission in the style of LISA, but with much smaller arms (1000 km) and much improved sensitivity at decihertz frequencies owing to its Fabry-Perot interfometers. We note that the high-frequency cutoff for our neutrino and matter components is simply the Nyquist frequency due to our data sampling. The lower frequency cutoff for the neutrino component is due to the length of our simulations $-$ our 3D models were carried out at most to $\sim$one second after bounce. Longer simulations will populate this lower frequency detection space. Third generation gravitational wave detectors will provide broadband coverage of galactic supernovae sensitive to both matter and neutrino asymmetries.

\section{Conclusions}\label{sec:conc}
Recent proliferation in 3D simulation capabilities of CCSNe in the first second after bounce have additionally provided new insights into the information contained in the gravitational waves sourced by matter and in neutrino signatures. However, study of gravitational waves sourced by neutrino asymmetries has lagged. Axisymmetric 2D studies overestimate the develop of axial instabilities and of the strains in general. Additionally, simplified 3D studies, often with parametrized neutrino heating, are insufficient to study the development of neutrino asymmetries critical to understanding neutrino gravitational waveforms. Thus, 3D simulations with detailed neutrino transport are essential to study with fidelity the neutrino waveforms for CCSNe. We presented here a study of the latter, which complements matter GWs and provides new insights. The neutrino component is fundamentally different from the matter component, involving variations of relativistic radiation on secular timescales. As a result, neutrino asymmetries of only a few percent can culminate in large gravitational strains.

Despite great advances in simulation studies of CCSNe this last decade, there is still much to accomplish. 
In particular, the gravitational wave emissions depend on the implementation of general relativity in a supernovae 
code, and in F{\sc{ornax}} we rely on a monopole approximation to relativistic gravity. In addition, improvements to the
radiation transport algorithm and the neutrino microphysics will alter the neutrino heating profile of CCSNe. 
Moreover, higher resolution studies (e.g. \citealt{nagakura_res}) may be required to capture the development of 
turbulence down to small scales and could result in modifications to the explosion and geometry. Lastly, longer-duration 
simulations, out to at least several seconds, are essential to understand the late-time asymptotic behavior of CCSNe.

In this paper, we looked at a sequence of progenitors from 9-60 M$_{\odot}$ evolved in 3D. Compared to the matter waveforms, we find that the neutrino waveforms can be up to two orders of magnitude larger in strain. Additionally, the neutrino waveform shows much less time variation and quasi-monotonic evolution due to the integral nature of the neutrino memory. However, whereas the matter component dominates at 100s to 1000s of Hertz, the neutrino component dominates at a few 10s of Hertz, and hence, contributes much less to the gravitational wave energy. Unlike the matter component, the neutrino contribution to the gravitational waveform does not have a prompt convective signal. However, both matter and neutrino GWs trace the development of turbulence in the first 100s of milliseconds after bounce. 

We find an approximate trend with progenitor mass and neutrino strain strength, which probes the strength of the accretion driving turbulence that manifests the GWs and identifies both small-scale instabilities, like convection, and larger-scale instabilities like the spiral SASI. The 9-M$_{\odot}$ progenitor, with a weak and short-lived accretion history, shows the smallest neutrino asymmetries and gravitational waveforms. However models, like the 25-M$_{\odot}$, which sustain longer and more powerful accretion experience larger sustained gravitational strains.

For models that do not explode, namely the 13-, 14-, and 15-M$_{\odot}$, we find signatures of the spiral SASI in both the neutrino luminosities and the neutrino gravitational waveforms at $\sim$100 Hz, coincident with the matter gravitational waveform \citep{vartanyan2019}. Due to the quadrupolar nature of the strain, its sign and magnitude illustrate the geometry of explosion and indicate, for instance, the development of equatorial or axial deformations in the propagating shock front. Additionally, the strain probes the precession of the spiral SASI for non-exploding models.

Lastly, we find that neutrino GWs can be detectable for galactic events by aLIGO as well as by DECIGO and the Einstein Telescope. Future detections of gravitational waves from galactic CCSNe will explore the development of turbulence, the explosion morphology, the accretion history and the success or failure of explosion. Additionally, the low frequency gravitational waves around one Hz, will probe the secular evolution of CCSNe, longer after explosion. CCSNe emit neutrinos for many seconds after explosion, and modeling these will require carrying out 3D simulations to much longer. 

\software{F{\sc{ornax}} \citep{skinner2019}}
 
\section*{Acknowledgments}

The authors acknowledge fruitful collaborations and discussions with David Radice, Hiroki Nagakura, Daniel Kasen, and Sherwood Richers.
DV and AB acknowledge support from the U.S. Department of Energy Office of Science and the Office
of Advanced Scientific Computing Research via the Scientific Discovery
through Advanced Computing (SciDAC4) program and Grant DE-SC0018297
(subaward 00009650) and support from the U.S. NSF under Grants AST-1714267
and PHY-1804048 (the latter via the Max-Planck/Princeton Center (MPPC) for Plasma Physics).
A generous award of computer time was provided
by the INCITE program. That research used resources of the
Argonne Leadership Computing Facility, which is a DOE Office of Science
User Facility supported under Contract DE-AC02-06CH11357. In addition, this overall research
project is part of the Blue Waters sustained-petascale computing project,
which is supported by the National Science Foundation (awards OCI-0725070
and ACI-1238993) and the state of Illinois. Blue Waters is a joint effort
of the University of Illinois at Urbana-Champaign and its National Center
for Supercomputing Applications. This general project is also part of
the ``Three-Dimensional Simulations of Core-Collapse Supernovae" PRAC
allocation support by the National Science Foundation (under award \#OAC-1809073).
Moreover, we acknowledge access under the local award \#TG-AST170045
to the resource Stampede2 in the Extreme Science and Engineering Discovery
Environment (XSEDE), which is supported by National Science Foundation grant
number ACI-1548562.  Finally, the authors employed computational resources provided by the TIGRESS high
performance computer center at Princeton University, which is jointly
supported by the Princeton Institute for Computational Science and
Engineering (PICSciE) and the Princeton University Office of Information
Technology, and acknowledge our continuing allocation at the National
Energy Research Scientific Computing Center (NERSC), which is
supported by the Office of Science of the US Department of Energy
(DOE) under contract DE-AC03-76SF00098.

\clearpage
\bibliographystyle{mnras}
\bibliography{References}

\end{document}